\def\etal{{\it et al.}}
\def\vs{{\it vs}}
\def\GeV{{\rm GeV}}

\def\bra#1{\left<#1\right|}
\def\ket#1{\left|#1\right>}
\def\vev#1{\left<#1\right>}
\def\simle{\mathrel{\rlap{\raise 0.511ex\hbox{$<$}}%
                         {\lower 0.511ex\hbox{$\sim$}}%
		}}
\def\simge{\mathrel{\rlap{\raise 0.511ex\hbox{$>$}}%
                         {\lower 0.511ex\hbox{$\sim$}}%
		}}
\def\TextStyleOver#1#2{%
	{\kern.05em\raise.3ex\hbox{\the\scriptfont0 $#1$}%
	\kern-.1em{\the\scriptfont0 /}%
	\kern-.1em\lower.2ex\hbox{\the\scriptfont0 $#2$}}}
\def\OneOver#1{\TextStyleOver{1}{#1}}
\def\BtoDstar{$B \to D^* \mu\,\nu$}
\def\CHat#1{\widehat{C}_{#1}}
\def\Set#1{${\cal #1}$}

\def\Eqn#1{Eqn.~\ref{eqn:#1}}
\def\Fig#1{Fig.~\ref{fig:#1}}
\def\Tab#1{Tab.~\ref{tab:#1}}

\def\Err(+#1-#2){\left(\vphantom{0}^{+#1}_{-#2}\right)}

\def\BorderBox#1#2{#2}

\def\InsertFigure[#1 #2 #3 #4]#5#6#7{
	\epsfxsize=#6%
	\epsfysize=#7%
	\epsfbox[#1 #2 #3 #4]{#5}%
	}

\def\BulletSect#1{\noindent$\bullet$\hskip0.30em{\bf #1}}

\def\PreprintNumber#1{\relax
{\normalsize\vskip-18mm\raggedright #1\vskip-\baselineskip\vskip18mm}}

\def\mytitle#1{\protect\title{
\PreprintNumber{Edinburgh 93/539, ~FERMILAB-CONF-93/390-T, ~hep-lat/9312088}%
#1}}

\documentstyle[epsf,twoside,fleqn,espcrc2]{article}

\mytitle{
The Isgur-Wise Function: A Lattice Determination from
Pseudoscalar~$\to$~Pseudoscalar Form Factors}

\author{
{\it UKQCD Collaboration} -- presented by James N. Simone\thanks{
present address:
Fermilab,
Box~500,
Batavia, IL,
60510, USA.
e-mail: simone@fnal.fnal.gov.
Talk given at
{\em Lattice '93}, Dallas, Texas, Oct.\ 12--16, 1993.
} \\
Department of Physics, University of Edinburgh, Edinburgh EH9~3JZ, Scotland
}

\begin{document}

\begin{abstract}
Form factors for pseudoscalar~$\to$~pseudoscalar decays of heavy-light
mesons are  found in quenched lattice QCD  with heavy-quark masses  in
the  range   of approximately   1-2  \GeV.   The Isgur-Wise  function,
$\xi(\omega)$, is extracted from these  form  factors.  Results are in
good  agreement with $\xi(\omega)$ derived  from CLEO measurements for
\BtoDstar.
\end{abstract}

\maketitle

\section{THE ISGUR-WISE FUNCTION}

Matrix elements are parameterized in terms of two form factors $h_\pm$
\begin{eqnarray}
\lefteqn{\frac{\bra{B(\vec{p}_b)}V^\mu\ket{A(\vec{p}_a)}}
     {\sqrt{m_a m_b}}
=h_+\left(\omega;m_a,m_b\right)(v_a + v_b)^\mu
+
} & & \nonumber \\
& & \quad\quad\quad\quad\quad h_-\left(\omega;m_a,m_b\right)(v_a - v_b)^\mu
\label{eqn:hFormFactDef}
\end{eqnarray}
where $v_{a}$ and $v_{b}$ are the meson four-velocities
and $\omega=v_a\cdot v_b$.

In  the heavy quark  limit, $m_{Q_{a,b}}\to\infty$,  form factor $h_-$
tends to zero   while $h_+$  approaches  $\xi(\omega)$, the  universal
Isgur-Wise form factor\cite{IsgurWise90}.

At finite heavy-quark mass, $h_\pm$ are still related to $\xi(\omega)$
although there  are now  both short-distance perturbative  corrections
and    nonperturbative   corrections  in powers    of   \OneOver{m_Q}.
Neglecting the power law corrections,
\begin{eqnarray}
h_+(\omega) &=
&\left[\CHat{1} + \frac{\omega+1}{2}
\left(\CHat{2} + \CHat{3}\right)\right]\xi_{\rm ren}(\omega)
\label{eqn:HplusToIW} \\
h_-(\omega) &=
&\frac{\omega+1}{2}\left[\CHat{2} - \CHat{3}\right]\xi_{\rm ren}(\omega)
\label{eqn:HminusToIW}
\end{eqnarray}
where the  Wilson  coeffieients $\CHat{i}$   have  been  computed   at
next-to-leading order by Neubert\cite{Neubert93}.

\section{METHODOLOGY}

An  $O(a)$-improved  fermion action\cite{SW85}  was  used to  generate
fermion  propagators for  60  quenched   gauge  configurations   on  a
$24^3\times48$  $\beta=6.2$ lattice\cite{UKQCD-Strange}.    The  three
light-quark masses,  $m_q$,  and the four  heavy-quark  masses, $m_Q$,
used here are also used in  our study of  $f_{D}$ and $f_{B}$ on these
same configurations\cite{Wittig93}.   Estimating the  heavy-quark mass
by the spin-average of the heavy-light pseudoscalar (P) and vector (V)
meson masses, in the $m_q\to0$ limit, we find, $m_Q\approx  1.5$, 1.9,
2.1, and  2.4 \GeV.  The light-quark  masses in ratio to strange quark
mass are $\TextStyleOver{m_q}{m_s}\approx 0.41$, 0.68, and 1.3.

We study euclidean three-point correlation functions
\begin{eqnarray}
\lefteqn{G^\mu(0,t,t_b;m_{Q_a},m_{Q_b},m_q,\vec{p}_a,\vec{p}_b)
=} & & \nonumber \\
& &\sum_{\vec{x},\vec{y}}
e^{i\vec{p}_b\cdot\vec{x}}
e^{i\vec{q}\cdot\vec{y}}
\vev{P_b(\vec{x},t_b)V^\mu(\vec{y},t)P_a^\dagger(\vec{0},0)}
\label{eqn:ThreePt}
\end{eqnarray}
where   $\vec{q}   =  \vec{p}_b -  \vec{p}_a$.  Operator $P_a^\dagger$
creates  a $Q_a\,\bar{q}$  pseudoscalar    and $P_b$  annihilates    a
$Q_b\,\bar{q}$   pseudoscalar.  The  current    $V^\mu$  is  a   local
$O(a)$-improved vector current\cite{Heatlie91}.

We  set $t_b=24$   and  symmetrize   correlators   about  this   time.
Correlators      have    lattice      momenta      $\vec{k}_b        =
({12a}/{\pi})\vec{p}_b=(0,0,0)$,          $(1,0,0)$,               and
$0\leq|\vec{k}_a|^2\leq2$. Quark mass $m_{Q_b}$ can be either $2.4$ or
$1.9\;\GeV$.

The ratio of a matrix element to the temporal component of the forward
matrix element of the flavor-conserving current is extracted by taking
the ratio of three-point functions
\begin{eqnarray}
\lefteqn{
         \frac{G^\mu(0,t,t_b;m_{Q_a},m_{Q_b},m_q,\vec{p}_a,\vec{p}_b)}
              {G^4  (0,t,t_b;m_{Q_b},m_{Q_b},m_q,\vec{p}_b,\vec{p}_b)}
         \stackrel{t_b\gg t\gg0}{\mbox{$-\!\!\!-\!\!\!-\!\!\!\longrightarrow$}}
        } & & \nonumber \\
& &
\frac{Z_a(\vec{p_a})}{Z_b(\vec{p_b})}
\frac{E_b}{E_a}\times
\frac{\bra{B(\vec{p}_b)}V^\mu\ket{A(\vec{p}_a)}}
     {\bra{B(\vec{p}_b)}V^4\ket{B(\vec{p}_b)}}
e^{-\delta\!E\, t}.
\label{eqn:ThreePtRatio}
\end{eqnarray}
For all Lorentz  components in the  ratio that are  non-zero, a single
minimal $\chi^2$ fit, using the full correlation  matrix, is found for
the  $t$  dependence  in  \Eqn{ThreePtRatio}.   Field   normalizations
$Z_{a,b}$, energies $E_{a,b}$, and $\delta\!E=E_a-E_b$ are constrained
to values obtained in fits   to the meson  propagators\cite{Wittig93}.
Equation~\ref{eqn:hFormFactDef}  is used  with  \Eqn{ThreePtRatio}  to
find     $h_\pm\left(\omega;m_a,m_b\right)/h_+\left(1;m_b,m_b\right)$.
After extrapolating $h_\pm$ to the $m_q\to0$ limit, relation
\Eqn{HplusToIW}     is   used to    extract  the  Isgur-Wise  function
$\xi(\omega)$ from $h_+(\omega)$.

For  flavor-conserving matrix elements,  $h_-$ should be exactly zero.
To test   this, we allow  both $h_\pm$  to be free   parameters in the
$\chi^2$   fit.   For   $m_{Q}=1.5\,\GeV$ and     $m_q\to0$  we   find
$|h_-|\simle0.1$ which is within $1\sigma$ of zero.  We then constrain
$h_-$ to zero in fits for flavor-conserving matrix elements.

\section{RESULTS}

\begin{table}[tb]
\renewcommand{\arraystretch}{1.30}
\caption{$\rho^2$ \vs\/ $\xi(\omega)$ model for parameter set \Set{A}.}
\label{tab:RhoSqSetA}
\begin{tabular}{@{}llll@{}} \hline
BSW		&pole		&ISGW 	&linear \\ \hline
$0.92\Err(+20-18)$ &$0.89\Err(+19-17)$ &$0.83\Err(+13-13)$ &$0.73\Err(+16-14)$
\\ \hline
\end{tabular}
\end{table}
\BulletSect{Slope Parameter}
The  slope  parameter,  $\rho^2\equiv-\xi^\prime(1)$,  is extracted by
finding    a minimum $\chi^2$  fit of    the  lattice $\xi(\omega)$ to
some
possible forms for the Isgur-Wise function
\begin{eqnarray}
\xi_{BSW}(\omega) &\!\!\!\!\!=
&\!\!\!\!\!\frac{2}{\omega + 1}
\exp\left(\left(1-2\rho^2_{BSW}\right)\frac{\omega-1}{\omega+1}\right)
\label{eqn:IW-BSW} \\
\xi_{pole}(\omega) &\!\!\!\!\!=
&\!\!\!\!\!\left(\frac{2}{\omega + 1}\right)^{2\rho_{pole}^2}
\label{eqn:IW-pole} \\
\xi_{ISGW}(\omega) &\!\!\!\!\!=
&\!\!\!\exp\left(-\rho_{ISGW}^2\left(\omega-1\right)\right)
\label{eqn:IW-ISGW}
\end{eqnarray}
as discussed in References~\cite{Neubert92}, \cite{Neubert93}, and
\cite{ISGW89} respectively.
Values obtained for $\rho^2$  should be  relatively insensitive to the
choice          of                     parameterization          since
Equations~\ref{eqn:IW-BSW}-\ref{eqn:IW-ISGW}   differ    only       at
$O\left((\omega-1)^2\right)$.

In  the  continuum  limit,   the   forward   matrix element  of    the
flavor-conserving vector current  has a  known normalization.   On the
lattice, matrix        elements        are            normalized    by
$\bra{B(\vec{p}_b)}V^4\ket{B(\vec{p}_b)}$  to reduce lattice artifacts
and to cancel the  local vector current  renormalization $Z_V$.  It is
important to  test the consistency  of this  normalization method.  We
fit lattice form  factors  to the function $N\,\xi_{BSW}(\omega)$ with
both $\rho^2$ and  the normalization, $N$,  determined by the $\chi^2$
fit.  Typically, $N$ differs from one by  $\simle3\%$  which is within
$1\sigma$.  We then constrained $N$ to one when finding $\rho^2$.

Label  as mass  set   \Set{A}\   the  combination of   quark   masses:
$m_{Q_b}=2.4\,\GeV$,  $m_{Q_a}={\rm  any\ }m_Q$    and  $m_q/m_s\to0$.
Values for   $\rho^2$  obtained    for  this   set   of  masses    and
Equations~\ref{eqn:IW-BSW}-\ref{eqn:IW-ISGW} are shown in
\Tab{RhoSqSetA}.    The      table     also     shows $\rho_{linear}^2$
from    $\xi_{linear}=1-\rho_{linear}^2(\omega-1)$.        Uncertainty
estimates are obtained by a bootstrap  procedure with only statistical
uncertainties  shown.   The     values  obtained  agree   with   other
determinations\cite{Shen93}             and          our       earlier
results\cite{UKQCD-iw-lett}.
\begin{figure}[tb]
\BorderBox{1pt}{%
\InsertFigure[40 20 490 480]{BtoDstar.ps}{212bp}{212bp}%
}\vskip-10.0mm
\caption{The quantity $V_{cb}\,\xi(\omega)$ measured by CLEO
for \protect\BtoDstar\  (diamonds). The
lattice form factor
(fancy squares) has been scaled by $|V_{cb}| = 0.034$ as
described in the text.
\label{fig:BtoDstar}}
\end{figure}

\BulletSect{Measured Form Factors}
In \Fig{BtoDstar}\  we compare the  lattice form factor  for  mass set
\Set{A}\ with  $|V_{cb}|\,\xi(\omega)$ derived from CLEO\cite{CLEO93}
data   for    \BtoDstar.     A      fit     of  the  CLEO  data     to
$|V_{cb}|\,\xi_{BSW}(\omega)$ with $\rho^2$ constrained to the lattice
value $\rho^2_{BSW}$ of
\Tab{RhoSqSetA} yields
\begin{equation}
|V_{cb}| = 0.034\Err(+3-3)\Err(+2-2)\,\sqrt{\frac{\tau_{B}}{1.49\,{\rm ps}}}.
\end{equation}
The  first   error is the  $\Delta\chi^2=1$  error in  the  fit to the
experimental  data and the  second   error reflects the uncertainty in
$\rho^2_{BSW}$.  Statistical uncertainties  in the lattice form factor
are of the same size as the errors in the experimental form factor.

The figure shows the lattice   $\xi(\omega)$ from ${\rm P}\to{\rm  P}$
transitions and $\xi(\omega)$ from CLEO ${\rm P}\to{\rm V}$ decay data
to  be remarkably similar  in shape.  Further  studies  of heavy quark
spin  symmetry using ${\rm   P}\to{\rm V}$  three-point  functions are
underway\cite{Hazel93,Hoeber93}.

\BulletSect{Heavy-Quark Mass Dependence}
In \Tab{RhoVSmQ} are values for $\rho^2_{BSW}$ from
separate analyses of
flavor-conserving matrix elements
with $m_Q=1.9$ and $2.4\,\GeV$.
\begin{table}[tb]
\renewcommand{\arraystretch}{1.30}
\caption{$\rho^2_{BSW}$ \vs\/ ${m_Q}\,(\GeV)$.}
\label{tab:RhoVSmQ}
\begin{tabular}{@{}lll@{}} \hline
$m_Q$		&$1.9$	&$2.4$ \\
$\rho^2$	&$0.91\Err(+43-20)$ &$1.06\Err(+66-34)$ \\ \hline
\end{tabular}
\end{table}
The errors  are large and the change  in $\rho^2$   with $m_Q$ is only
about $0.5\sigma$ over the range of $m_Q$ studied.

The $O(\OneOver{m_Q})$  corrections  to \Eqn{HplusToIW}\ that  relates
$h_+(\omega)$ to $\xi(\omega)$ may be small  since,
by Luke's theorem\cite{Luke90},
there can be at most $O(\OneOver{m_Q^2})$ corrections to this relation
at  $\omega=1$.

For  mass   set  \Set{A}   with $|\vec{k}_b|=0$, the
variations in the values  of $\xi(1)$  extracted from $h_+(1;m_a,m_b)$
are $\simle0.5\%$ as $m_{Q_a}$ is varied over the four possible values
of $m_{Q}$.    The  differences  are   smaller  than the   statistical
uncertainties.  For $|\vec{k}_b|=1$, the variations in $\xi(1)$ values
are now as much as ten times  larger  than for the zero momentum case.
However, the differences are still within $1\sigma$ of zero.

Tests using  the relation in  \Eqn{HminusToIW}, which is not protected
from  $O(\OneOver{m_Q})$ corrections   by  Luke's theorem,  are   more
sensitive indicators of $m_Q$ effects\cite{Henty93}.  A study of $h_-$
may then  help characterize the nonperturbative  power law corrections
to $\xi(\omega)$ at finite $m_{Q}$.

\BulletSect{Light-Quark Mass Dependence}
In \Tab{RhoVSmq} we show $\rho^2_{BSW}$ at fixed light-quark mass
for the heavy-quark masses of set \Set{A}.
These values should also be compared the value of $\rho^2_{BSW}$ in
\Tab{RhoSqSetA} obtained in the chiral limit.
The trend is for $\rho^2$ to decrease with decreasing light-quark
mass.
Further work is necessary to understand the chiral behavior of $\xi(\omega)$.
\begin{table}[t]
\renewcommand{\arraystretch}{1.30}
\caption{$\rho^2_{BSW}$ \vs\/ \TextStyleOver{m_q}{m_s}.}
\label{tab:RhoVSmq}
\begin{tabular}{@{}llll@{}} \hline
\TextStyleOver{m_q}{m_s}	&$0.41$	&$0.68$ &$1.3~$\\
$\rho^2$	&$1.09\Err(+24-11)$ &$1.19\Err(+17-10)$	&$1.31\Err(+15-~6)$
\\ \hline
\end{tabular}
\end{table}

\section{CONCLUSION}

Using heavy quark symmetry and the lattice is an effective way to study
$B\to D$ decays.

\section*{ACKNOWLEDGEMENTS}

The authors wish to acknowledge their conversations with
C. Bernard, J. Mandula, M. Ogilvie, Y. Shen, and A. Soni
concerning this work.
This work was carried out on a Meiko i860
Computing Surface supported
by SERC Grant GR/32779, the
University of Edinburgh and Meiko Limited.

\end{document}